\documentclass[aps, preprint,amsmath,amssymb,eshowkeys,nofootinbib]{revtex4-1}

\usepackage{verbatim,graphics,graphicx,color,slashed,textcomp,bbm,mathdots,multirow,array}
\usepackage{dcolumn}
\usepackage{bm}
\usepackage{revsymb4-2}
\usepackage{autobreak}

\graphicspath{{figures/}}
\usepackage[colorlinks=true,linkcolor=red,urlcolor=blue,citecolor=blue]{hyperref}

\begin{document}

\title{Universal Density and Velocity Distributions of Dark Matter around Massive Black Holes}


\author{Zi-Chang Zhang$^{a}$} 

\author{Hai-Chao Yuan$^{a}$} 

\author{Yong Tang$^{a,b,c}$}
\affiliation{\begin{footnotesize}
		${}^a$School of Astronomy and Space Science, University of Chinese Academy of Sciences (UCAS), Beijing 100049, China\\
		${}^b$School of Fundamental Physics and Mathematical Sciences, \\
		Hangzhou Institute for Advanced Study, UCAS, Hangzhou 310024, China \\
            ${}^c$International Center for Theoretical Physics Asia-Pacific, Beijing/Hangzhou, China 
		\end{footnotesize}}

\date{\today}

\begin{abstract}
The distribution of dark matter at the galactic center, crucial for indirect searches, remains uncertain. In particular, in the vicinity of the massive black hole in the center of a galaxy where indirect signals may be stronger, the density of a dark matter spike may undergo redistribution. Here we calculate the density surrounding Schwarzschild black holes that originate from diverse initial dark halos and estimate the velocity distribution of dark matter particles. By employing a series of Hernquist and power-law initial dark halos, we obtain a fitting formula between dark matter spikes and black hole masses. The Maxwell-Boltzmann distribution is utilized to approximate the velocity distribution of dark matter particles. As an application, taking into account dark matter self-annihilation, we assess the relic densities of dark matter spikes around black holes. We find that the relic spikes for s-wave annihilation are higher than p-wave annihilation, and the relic densities obtained for p-wave annihilation depend on the velocity distribution, varying significantly with distance. The findings shall further provide useful insights for multi-messenger dark matter detections in the future. 

\end{abstract}


\maketitle


\section{Introduction} \label{sec:intro}
Dark matter plays an important role in the formation of galaxies and the evolution of the universe.
Numerous astronomical observations have provided compelling evidence for the existence of dark matter, including rotation curves of spiral galaxies, the colliding Bullet Cluster, gravitational lensing, large-scale structure, and anisotropies in the cosmic microwave background, among others. 
However, the nature of dark matter remains a mystery.
As a popular dark matter candidate, weakly-interacting massive particles (WIMPs) might self-annihilate or decay to standard model particles, 
producing gamma rays or cosmic rays, which could be detected in indirect detection experiments \cite{doi.10.1142/S0217751X13300408,BERTONE2005279,DONATO201441}.
To indirectly detect dark matter at an astronomical scale, a concrete distribution of dark matter in galaxies is necessary. 
There are various classical models, including the power-law profile, Einasto profile \cite{1965TrAlm...5...87E}, Hernquist profile \cite{1990ApJ...356..359H}, and Navarro-Frenk-White (NFW) profile \cite{Navarro_1997}, which effectively describe different aspects of the external structure of a dark halo. However, the distribution of dark matter in galactic centers where the indirect signals might be stronger, is still uncertain. 

The profile of dark matter can affect many astrophysical and astronomical observations. For example, dark matter density has a direct impact on the observations of gamma-ray flux in the galactic center \cite{BERGSTROM1998137, buckley2013cosmicfrontierindirectdark, Ackermann_2017, Lacroix_2018} and substructure of halo~\cite{10.1111/j.1365-2966.2007.12828.x,Xie:2025udx}. 
And the annihilation cross section in dark matter mini-spikes around intermediate-mass black holes might vary with astrophysical parameters \cite{PhysRevD.72.103517}. 
In recent years, gravitational waves have been discussed as a novel indirect method for probing dark matter~\cite{Cole2023DistinguishingEE,PhysRevD.102.083006, PhysRevLett.110.221101, 10.21468/SciPostPhysCore.3.2.007,PhysRevD.105.043009}.  Typical gravitational wave sources for searching dark matter are intermediate-mass-ratio inspirals (IMRIs) and extreme-mass-ratio inspirals (EMRIs), which are the main targets of the future space-based gravitational-wave observatories like LISA \cite{2017arXiv170200786A,baker2019laserinterferometerspaceantenna}, Taiji \cite{10.1093/nsr/nwx116, 2020NatAs...4..108R} and TianQin \cite{Luo_2016}. These binaries consist of a supermassive black hole and a smaller compact object. 
As an environmental effect, dark matter influences the evolutions of IMRIs and EMRIs through dynamical friction~\cite{1943ApJ....97..255C}, which can be indirectly detected in orbital evolution \cite{Antonini_2012,Chan_2023,Chan_2024} and gravitational wave signal \cite{Yue_2019,Fakhry_2023}. 
Recent studies on the orbit of S-star in the Milky Way Galaxy have limited the mass of dark matter surrounded by orbit to be less than $1200M_\odot$ \cite{Nampalliwar_2021, Shen:2023kkm, Will_2023, 2024arXiv240912261T}. 
In detail, the velocity distribution of dark matter has been considered in the calculations of annihilation cross section \cite{PhysRevD.87.115007,PhysRevD.79.083525,xu2024directdetectionhiggsportal,phoroutanmehr2024relaxingconstraintsdarkmatter} 
and dynamical friction \cite{PhysRevD.105.063029, 2022SCPMA..6500412L}. The potential impact of dark matter on astrophysical and astronomical observation is becoming increasingly important. 
But the velocity distribution of dark matter in relativistic spike close to a black hole is rarely considered. 

The profile of the inner density of dark matter is usually associated with the supermassive black hole at the center of the galaxy. 
Dark matter spike is frequently-used to describe dark matter near black holes. 
Ref.~\cite{PhysRevLett.83.1719} discussed that under the influence of the black hole adiabatic growth, the dark matter at the center of the galaxy might redistribute into a spike, which is higher than initial profiles. And they also consider that dark matter annihilation shall decrease the spike to obtain a relic density of dark matter spike.
Adiabatic growth with different initial distribution functions was investigated~\cite{MacMillan_2002}, and various resulting spikes were obtained. 
In the vicinity of inactive supermassive black holes at galactic center, dark matter spike can maintain prolonged stability~\cite{PhysRevD.64.043504}. 
Additional widely considered astrophysical processes might change the profile of the spike, including stellar scattering~\cite{PhysRevLett.92.201304,PhysRevLett.93.061302}, black hole merger~\cite{PhysRevLett.88.191301,PhysRevD.72.103502}, self-interaction of dark matter~\cite{PhysRevD.89.023506}, which stimulated studies such as the evolution of dark matter spike~\cite{PhysRevD.78.083506,Eroshenko_2024}, 
observations~\cite{Bromley_2011,PhysRevD.89.063534,Schnittman_2015}, 
the effects on black holes
~\cite{ZHAO2002385,Xu_2021,doi:10.1142/S0217751X05020756} and so on.
Moreover, fully relativistic analysis around a Schwarzschild black hole \cite{PhysRevD.88.063522} predicted a higher density of spike than in the Newtonian case. 
Furthermore, in the environment of a rotating black hole, the profile of the spike varies with spin factor and the angle with respect to the black hole rotation axis \cite{PhysRevD.96.083014}. 
More recently, ‘dark matter mounds’ \cite{bertone2024darkmattermoundsrealistic} demonstrated a dark matter overdensity, which is time-dependent and shallower than dark matter spikes. 
Ref.~\cite{maeda2024einsteinclustercentralspiky} found that dark matter in the range of $3GM < r< 6GM$ is metastable by Einstein cluster construction.  
The spikes that grow from NFW profile and Hernquist profile with different initial parameters (the total mass and distance scale of dark halo) were calculated by \cite{PhysRevD.106.044027} and \cite{chakraborty2024tidallovenumbersquasinormal}. However, a systematic investigation into the correlation between the parameters of the dark halo and the mass of the black hole remains lacking.

In this paper, we focus on the inner environment of a galaxy and systematically investigate the correlation between the parameters of the dark halo and the mass of the black hole, substantially extending our previous work~\cite{PhysRevD.110.103008}. Around Schwarzschild black holes, we use relativistic analysis to calculate dark matter spikes that grow from Hernquist halos and power-law halos with different initial parameters. The parameters of the dark halo are related to the black hole mass by the mass-velocity-dispersion relation. We fit the spike characteristic density to the black hole mass and get a logarithmic relation.  
We analyze the velocity range of dark matter particles in the spike, and fit the velocity distribution  that can be useful for further numerical analysis. Then, we consider the effect of dark matter annihilation on the density. We utilize the calculated velocity distribution to evaluate the velocity-averaged annihilation cross section of dark matter at different distances, and estimate the relic density of the spike. 
Two types of dark matter annihilation (s-wave and p-wave) exert different levels of influence on the relic density of the spike. 
We anticipate that it will provide more detailed insights in future astronomical observations.

The contents are organized as follows. In Sec.~\ref{sec:dark matter}, we outline the theoretical framework for our numerical analysis and establish the conventions. In Sec.~\ref{sec:density_profile} we calculate the dark matter spikes around Schwarzschild black holes with mass ranging between $10^4 M_\odot$ and $10^9 M_\odot$, employing both Hernquist and power-law profiles for comparison. And we fit the resulting density profiles with an analytical formula. In Sec.~\ref{sec:velocity}, we discuss the velocity distribution of dark matter close to the black hole. In Sec.~\ref{sec:anni}, we investigate the dark matter spike with annihilation and showcase the relic density profile. 
In Sec.~\ref{sec:discussion}, we compare the differences between our methods and the attractor solutions in~\cite{10.1111/j.13652966.2011.18687.x, 10.1111/j.13652966.2011.19258.x}.
Finally, we summarize in Sec.~\ref{sec:CON}. We use natural unit $c=1$ throughout this paper. 

\section{Theoretical Framework} \label{sec:dark matter}
In this section, we establish the theoretical framework and its conventions for the relativistic calculation of the dark matter density spike around an adiabatically growing black hole, closely following previous work~\cite{PhysRevD.88.063522} and~\cite{PhysRevD.110.103008}. 
Then we outline how we determine the physical parameters of dark halos for black holes in the mass range $10^4 M_\odot -10^9 M_\odot$. With the initial parameters and density profiles, we estimate dark matter spikes around Schwarzschild black holes by implementing numerical integration. Finally, we give the fitted relation between the parameters of the dark matter spike and the mass of the black hole.

We consider the case that dark matter particles around a black hole are collisionless. They follow a stationary distribution function $f^{(4)}(x,p)$, normalized by $\int f^{(4)}(x,p)\sqrt{-g} d^4p = 1$. 
At a specific position, the mass current four-vector is determined by
\begin{equation}\label{mass current}
J^\mu(x)=\int f^{(4)}(x,p)u^\mu \sqrt{-g} d^4p ,
\end{equation}
where $x$ is the position coordinate, $p$ is the four-momentum, $u^\mu = {p^\mu}{/}{\mu}$ is the four-velocity, 
$\mu$ is the rest mass of dark matter particle,  
$g$ is the determinant of the metric $g_{\mu\nu}$, and $d^4p$ is the four-momentum volume element. 
Using the relations that $J^\mu=\rho u^\mu$ and $u_\mu u^\mu=-1$,  we can express the density in a local free-falling frame as $\rho = \sqrt{-g_{\mu\nu}J^\mu J^\nu}$. 
See Appendix.~\ref{app:stability} for the justification of stable profile.

Changing the integration element $d^4p$ to four constants of motion $d\varepsilon dL^2dL_zd\mu$, we can rewrite the density of spike as~\cite{PhysRevD.88.063522, PhysRevD.110.103008}
\begin{equation}\label{density eqs}
\rho(r) = \sqrt{1-\frac{2GM}{r}}\frac{2}{r^2}\int d\varepsilon dL^2 dL_z \frac{\varepsilon f(\varepsilon,L^2,L_z)}{\sqrt{\varepsilon^2 - (1-{2GM}/{r})(1+{L^2}/{r^2})}\sqrt{L^2 - L_z^2}},
\end{equation}
where $\varepsilon$, $L$, and $L_z$ are the three constants of motion, relative energy, total angular momentum, and angular momentum along the z-direction, respectively. 
Here, the integration over $\mu$ has been performed by simplifying the distribution function as $f^{(4)}(x,p) = \mu^{-3}f(\varepsilon,L^2,L_z)\delta(\mu-\mu_0)$. Because we consider that the dark matter particles in spike have the same rest mass $\mu_0$.

To calculate the density $\rho(r)$, a concrete form of distribution function $f(\varepsilon,L^2,L_z)$ is necessary. 
However, the distribution function $f(\varepsilon,L^2,L_z)$ cannot be analyzed directly. 
Considering the adiabatic growth of black hole is a physically viable method to start. 
In the method a black hole is put into the galaxy center with preexisting initial dark halo and adiabatically grows. 
In this process, the time scale of growth is longer than the orbital evolution of dark matter particles. 
The initial halo shall be redistributed to the dark matter spike near the galactic center, and it retains the initial profile far away from black hole.

A following result for black hole adiabatic growth is that the integrals of the motion $I_{r,\theta,\phi}$ and the distribution function $f(\varepsilon,L^2,L_z)$ are invariant~\cite{Merritt+2013,PhysRevD.96.083014}. 
Consequently, we can trace the initial distribution function by $I^{'}_{r,\theta,\phi}= I_{r,\theta,\phi}$ and $f^{'}(\varepsilon^{'},L^{'2},L^{'}_z)=f(\varepsilon,L^2,L_z)$, 
thereby numerically implementing the calculation of Eq.~\ref{density eqs}. 
In this method, only dark matter particles on bounded orbits contribute to the density at specific distance $r$, and unbounded dark matter particles have escaped or fallen into the black hole, so that $\rho(r)$ might be an optimistic lower limit estimation. 
Furthermore, we consider the case that the elapsed time is long enough to ensure that the relaxation time scale ($10^8-10^9$yr) of dark matter particles and the growth time scale ($10^9-10^{10}$yr) of black holes are met. 
Otherwise, dark matter spike might not form.
The astrophysical processes mentioned in Sec.~\ref{sec:intro} may disrupt the spikes. Therefore, for simplicity, we first consider that the central black holes have avoided major mergers and the dark matter is not self-interacting to sustain the stability of spike, which implies that the massive black hole is inactive.

Now we need to set the initial dark halos to compute the density. Given its $r^{-1}$ behavior near the galactic center, similar to the NFW profile, but with better convergence properties, we adopt the Hernquist profile as the initial dark halo model \cite{1990ApJ...356..359H},
\begin{equation}\label{Hernquist profile}
\rho_H(r) = \frac{\rho_0}{(r/r_s)(1+r/r_s)^3},
\end{equation}
where $\rho_0$ is characteristic density, $r_s$ is the characteristic scale of halo, and the total mass of dark halo $M_\text{halo} = 2\pi \rho_0 r_s^3$.
The phase space distribution function of Hernquist halo can be analytically expressed by Eddington inversion method \cite{1990ApJ...356..359H,Binney2008GalacticDS,Lacroix_2018},
\begin{equation}
f_H(\tilde{\epsilon}) = \frac{M_\text{halo}}{\sqrt{2}(2\pi)^3(GM_\text{halo}r_s)^{3/2}}\frac{\sqrt{\tilde{\epsilon}}}{(1 - \tilde{\epsilon})^2} \left[ (1-2\tilde{\epsilon})(8{\tilde{\epsilon}}^2 - 2\tilde{\epsilon} - 3) + \frac{3\sin^{-1}{\sqrt{\tilde{\epsilon}}}}{\sqrt{\tilde{\epsilon}(1-\tilde{\epsilon})}}  \right].
\end{equation}
Here $\tilde{\epsilon} = -r_s E^{\prime}/GM_\text{halo}$ is dimensionless relative energy and $E^{\prime}$ is the relative energy of per unit particle mass in Newtonian case.

Under the condition of adiabatic growth, the distribution function $f(\varepsilon,L^2,L_z)$ of dark matter spike can be computed by initial distribution function $f_H(\tilde{\epsilon})$ before adiabatic growth. And the initial and final constants of motion are related by the adiabatic invariants $I_{r,\theta,\phi}$ of the halo before and after adiabatic growth. The subscripts $r,\theta$ and $\phi$ represent the radial, azimuthal, and polar components of invariants, respectively.
The adiabatic invariants of dark matter in the Hernquist halo are
\begin{equation}\label{IH}
\begin{aligned}
I_r^{H}(E^{\prime},L^{\prime}) &=\oint dr \sqrt{2E^{\prime} - 2\Phi_H -\frac{{L^{\prime}}^2}{r^2}},\\
I_\theta^H(L^{\prime},L_z^{\prime})&=2\pi \left| L^{\prime} - L_z^{\prime}\right|,\\
I_\phi^H(L_z^{\prime})&=2\pi   L_z,
\end{aligned}
\end{equation}
where 
\begin{equation}
\Phi_H = -\frac{GM_\text{halo}}{r + r_s}.
\end{equation}
The invariants after growth in Schwarzschild background geometry are
\begin{equation}\label{IS}
\begin{aligned}
I_r^S(\varepsilon,L^2,L_z)&=\oint dr \frac{\sqrt{\varepsilon^2 - (1-{2GM}/{r})(1+{L^2}/{r^2})}}{1-2GM/r},\\
I_\theta^S(L,L_z)&=2\pi \left| L - L_z\right|,\\
I_\phi^S(L_z)&=2\pi L_z.
\end{aligned}
\end{equation}
Matching $I_{r,\theta,\phi}^S = I_{r,\theta,\phi}^H$, the values of distribution function $f$ in Eq.~\ref{density eqs} are determined,  
and the integration can be evaluated by the Monte Carlo method. 
As we calculate the density $\rho(r)$ at $r$, we randomly sample the variables $(\varepsilon,L,L_z)$ in the following regions, 
\begin{equation}\label{boundary}
\begin{aligned}
\varepsilon &\in  [\varepsilon_\text{min}(r),1],\\
L^2 &\in [L^2_\text{crit}(r), L^2_\text{max}(r)],\\
L_z &\in [-|L|,|L|]. 
\end{aligned}
\end{equation}
Here the extrema $\varepsilon_\text{min}$, $L_\text{max}$ and $L_\text{crit}$ are obtained by black hole capture condition ~\cite{PhysRevD.88.063522,PhysRevD.96.083014}. 
Then Eq.~\ref{density eqs} can be evaluated by 
\begin{equation}\label{density_num}
\begin{aligned}
\rho(r) = \sqrt{1-\frac{2GM}{r}}\frac{2}{r^2}  \sum_{i=1}^n 
\frac{2|L_i|[1-\varepsilon_\text{min}(r)][L^2_\text{max}(r)-L^2_\text{crit}(r)]\varepsilon_i f(\varepsilon_i,L^2_i,(L_z)_i)}{N\sqrt{\varepsilon_i^2 - (1-{2GM}/{r})(1+{L_i^2}/{r^2})}\sqrt{L^2_i - (L_z^2)_i}},
\end{aligned}
\end{equation}
where $n$ is the number of points that satisfy $I_{r,\theta,\phi}^S = I_{r,\theta,\phi}^H$ and the expressions inside the square roots of Eq.~\ref{density_num} $>0$, and $N$ is the number of total sampling points. 

Subsequently, we examine the correlation between dark matter spikes and black holes of different masses. 
The black hole mass $M$ and the parameters of the initial dark halo ($M_\text{halo}, r_s$) are related by the mass-velocity-dispersion relation, namely the one-dimensional velocity dispersion of halo and the virial mass relations \cite{Ferrarese_2000,Gebhardt_2000,PhysRevD.102.103022, PhysRevD.99.043533},
\begin{equation}\label{M-v}
\begin{aligned}
\rm{log} (\mathit{M/M_\odot}) &=8.12+4.24\; \rm{log}(\sigma/200 \rm{km\cdot s^{-1}}),\\
M_\text{vir}&=4\pi \rho_0 r_s^3 g(c(M_\text{vir}))=\frac{4}{3}\pi R_\text{vir}^3 \times 200\rho_\text{crit},\\
\sigma^2 &=\frac{4\pi G \rho_0 r_s^2 g(c_\text{m})}{c_\text{m}},
\end{aligned}
\end{equation}
where $\sigma$ is the one-dimensinal velocity dispersion of the halo, $M_\text{vir}$ is the virial mass, $R_\text{vir}$ is the virial radius, $c(M_\text{vir})=r/r_s$ is the characteristic distance ratio, $g(x)$ is the mass integral function of the halo, $c_\text{m}$ the distance maximizes the circular velocity $\sqrt{GM(r)/r}$, and $\rho_\text{crit}$ is the critical density. 
Here we take $g(x)=x^2/2(x+1)^2$, which is the mass function of the Hernquist profile, and accordingly $c_\text{m}=1$. 
The concentration relation of the virial distance ratio $c(M_\text{vir}) = R_\text{vir}/r_s$ is given by \cite{10.1093/mnras/stu1014} 
\begin{equation}\label{concentration}
c(M_\text{vir}) = \sum_{i=0}^5 c_i \log^i\left(\frac{h_0 M_\text{vir}}{M_\odot}\right),
\end{equation}
where $h_0=0.67$ and $c_i(i=0,1,2,3,4,5)=(37.5153, -1.5093, 1.636\times 10^{-2}, 3.66\times 10^{-4}, -2.89237\times 10^{-5}, 5.32\times 10^{-7})$. 
Then for a given black hole mass, we can obtain the parameters $(M_\text{halo},r_s)$ of the dark halo. In Table.~\ref{tab:halo parameter} we tabulate the parameters of the initial dark halo for black hole mass ranging from $10^4 M_\odot$ to $10^9 M_\odot$, which shall serve as the inputs for our later investigation of dark matter spikes. 
For further comparison, we shall also calculate the spikes growing from power-law profiles 
$\rho(r)=\rho_0(r_s/r)^\gamma$ by following~\cite{PhysRevLett.83.1719}. Considering that the halo mass of the power-law profile is divergent as $r\rightarrow\infty$, we set the halo masses consistent with Hernquist profiles as $r=r_s$, and characteristic density $\rho_0$ is also the same as Hernqusit halo. 
Then the characteristic scale $r_s$ of power-law profiles are determined by $M_{halo} = 4\pi \rho_0 r_s^3/(3-\alpha)$. 
We show the initial parameters of power-law halos in Tabel.~\ref{tab:halo parameter power law}. 

\begin{table*}[t]
\caption{\label{tab:halo parameter}%
The parameters ($M_\text{halo}, r_s$) of the initial Hernquist halo surrounding black holes with mass range from $10^4 M_\odot$ to $10^9 M_\odot$.}
\begin{ruledtabular}
\begin{tabular}{ccc}
Black hole mass ($M/M_\odot$)&Halo mass ($M_\text{halo}/M_\odot$)& Characteristic scale of halo ($r_s/\text{kpc}$)\\ \hline
 $1\times10^4 $ & $4.64\times10^8 $& $1.09$  \\
 $5\times10^4 $ & $1.63\times10^9 $& $1.79$ \\
 $1\times10^5 $ & $2.80\times10^9 $& $2.22$ \\
 $5\times10^5 $ & $9.95\times10^9 $& $3.69$ \\
 $1\times10^6 $ & $1.72\times10^{10} $& $4.59$ \\
 $5\times10^6 $ & $6.16\times10^{10} $& $7.70$ \\
 $1\times10^7 $ & $1.06\times10^{11} $& $9.63$ \\
 $5\times10^7 $ & $3.84\times10^{11} $& $16.3$ \\
 $1\times10^8 $ & $6.69\times10^{11} $& $20.4$ \\
 $5\times10^8 $ & $2.43\times10^{12} $& $34.6$ \\
 $1\times10^9 $ & $4.22\times10^{12} $& $43.4$ \\
\end{tabular}
\end{ruledtabular}
\end{table*}

\begin{table*}[ht!]
\caption{\label{tab:halo parameter power law}%
The parameters ($M_\text{halo}, r_s, \gamma$) of initial power-law profiles surrounding black holes with mass range from $10^4 M_\odot$ to $10^9 M_\odot$.}
\begin{ruledtabular}
\begin{tabular}{ccccc}
 Black hole mass&\multicolumn{4}{c}{Initial parameters $\gamma$ and $(M_\text{halo}/M_\odot, r_s/\text{kpc})$} \\
 $M/M_\odot$ &$\gamma=1$ & $\gamma=1.25$ &$\gamma=1.5$ & $\gamma=1.75$\\ \hline
 $1\times10^4 $ & $(4.64\times10^8,1.09)$& $(4.64\times10^8,1.04)$ & $(4.64\times10^8,0.989)$& $(4.64\times10^8,0.930)$\\
 $1\times10^5 $ & $(2.80\times10^9,2.22)$& $(2.80\times10^9,2.12)$ & $(2.80\times10^9,2.02)$& $(2.80\times10^9,1.90)$\\
 $1\times10^6 $ & $(1.72\times10^{10},4.59)$& $(1.72\times10^{10},4.40)$ & $(1.72\times10^{10},4.17)$& $(1.72\times10^{10},3.93)$\\
 $1\times10^7 $ & $(1.06\times10^{11},9.61)$& $(1.06\times10^{11},9.20)$ & $(1.06\times10^{11},8.75)$& $(1.06\times10^{11},8.23)$\\
 $1\times10^8 $ & $(6.69\times10^{11},20.4)$& $(6.69\times10^{11},19.5)$ & $(6.69\times10^{11},18.5)$& $(6.69\times10^{11},17.4)$\\
 $1\times10^9 $ & $(4.22\times10^{12},43.4)$& $(4.22\times10^{12},41.5)$ & $(4.22\times10^{12},39.4)$& $(4.22\times10^{12},37.1)$\\
\end{tabular}
\end{ruledtabular}
\end{table*}

\section{Density Profile}\label{sec:density_profile}
\begin{figure*}[t]
\centering
\includegraphics[width=5in]{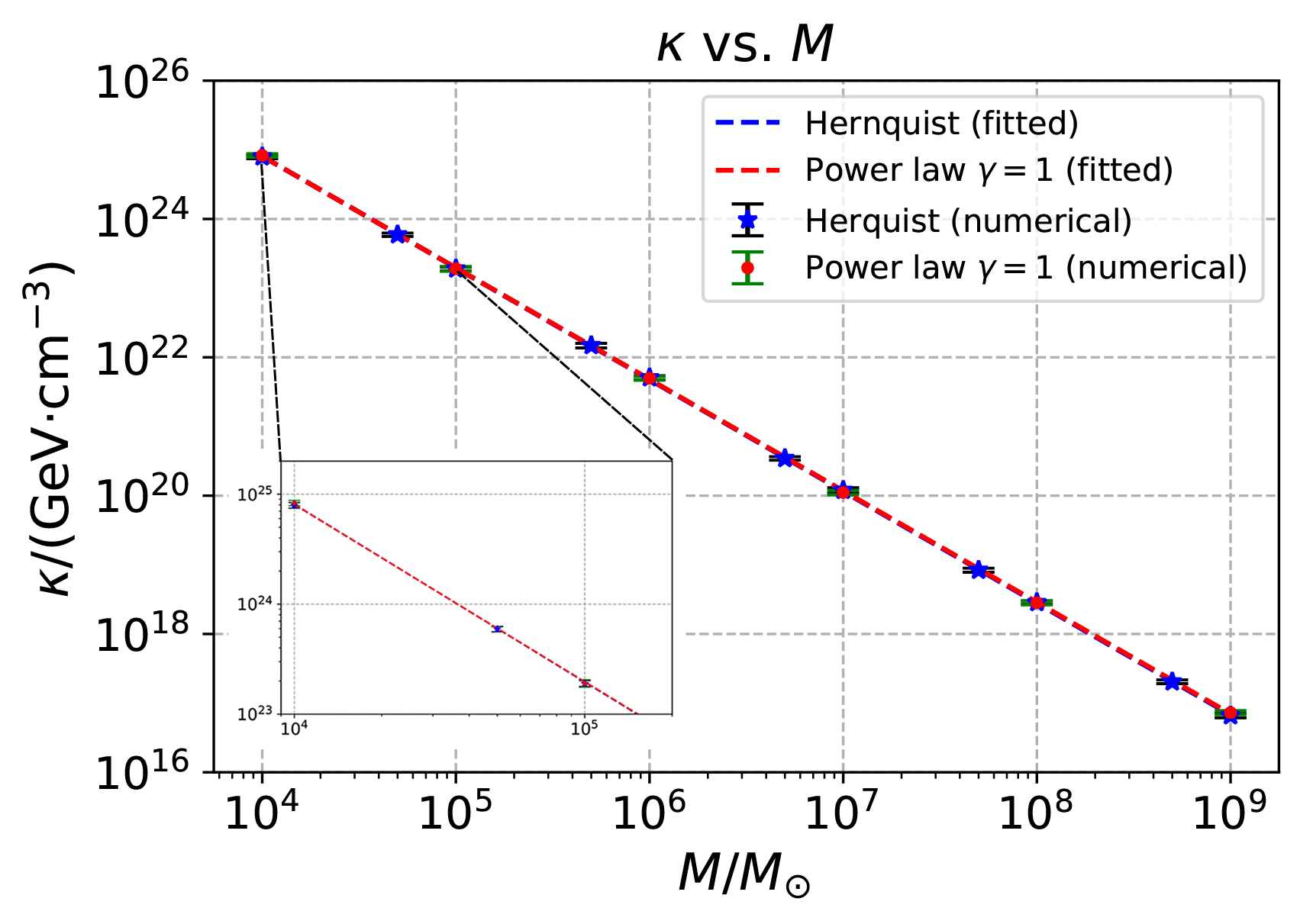}
\caption{The logarithmic relation between characteristic density of spike $\kappa$ and black hole mass $M$. The dashed lines are the fitting results following Eq.~\ref{lg-lg}. 
The blue dashed line is the characteristic density for the spike growing from the initial Hernquist halo, and the red dashed line is for the spike growing from the initial power-law profile with $\gamma=1$. The scattered data points with error bars represent numerical data points calculated based on the initial halo parameters listed in Table.~\ref{tab:halo parameter}.\label{kappa}}
\end{figure*}

\begin{figure*}[ht!]
\centering
\includegraphics[width=0.48\textwidth,height = 0.38\textwidth]{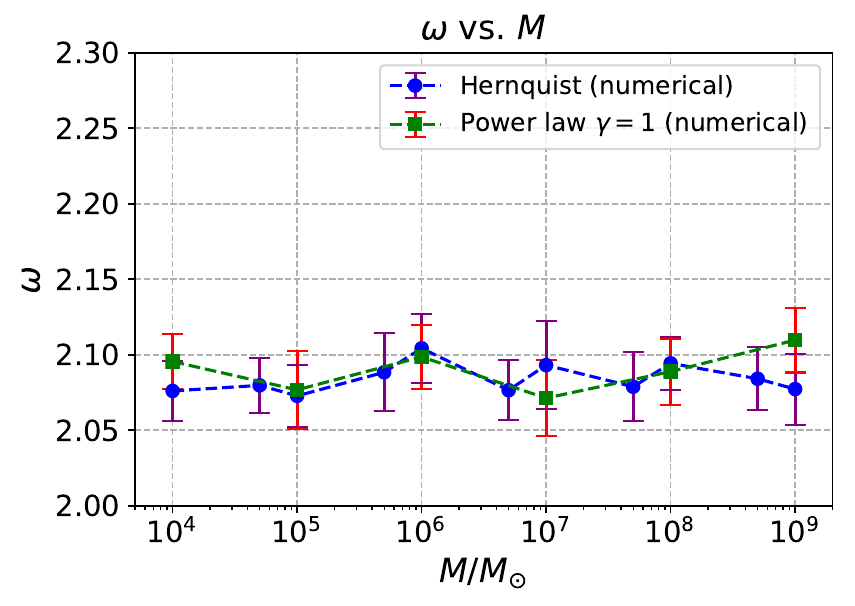}
\includegraphics[width=0.48\textwidth,height = 0.38\textwidth]{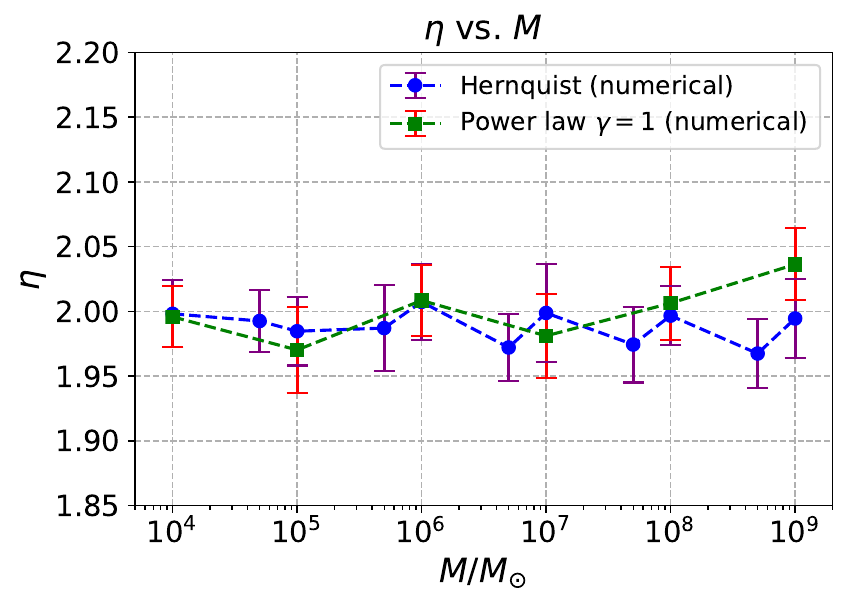}
\caption{The numerical scatters with error bar of $\omega$ and $\eta$ for spikes growing from initial Henrquist and power-law profiles with $\gamma=1$. The blue dashed line: Hernquist profile; the green dashed line: power-law profile with $\gamma=1$. (Left) Numerical scatters of $\omega$ vs. black hole mass $M$. (Right) Numerical scatters of $\eta$ vs. black hole mass $M$.\label{omega and eta}}
\end{figure*}

\begin{figure*}[t]
\centering
\includegraphics[width=5in]{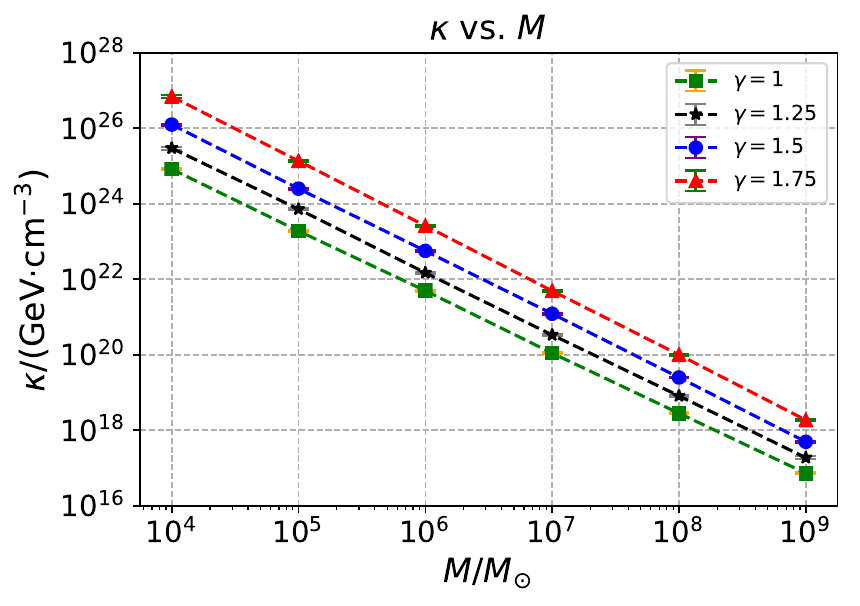}
\caption{The logarithmic relation between characteristic density of spike $\kappa$ and black hole mass $M$. The dashed lines are the fitting results following Eq.~\ref{lg-lg}. The green dashed line, blue dashed line, black dashed line,  and red dashed line are the results of power-law profiles with $\gamma=1,1.25,1.5$ and $1.75$ respectively. The scattered data points with error bars on homologous curves represent numerical data points calculated based on the initial halo parameters listed in Table.~\ref{tab:halo parameter}.\label{fig:kappa1}}
\end{figure*}

\begin{figure*}[t]
\centering
\includegraphics[width=0.48\textwidth,height = 0.38\textwidth]{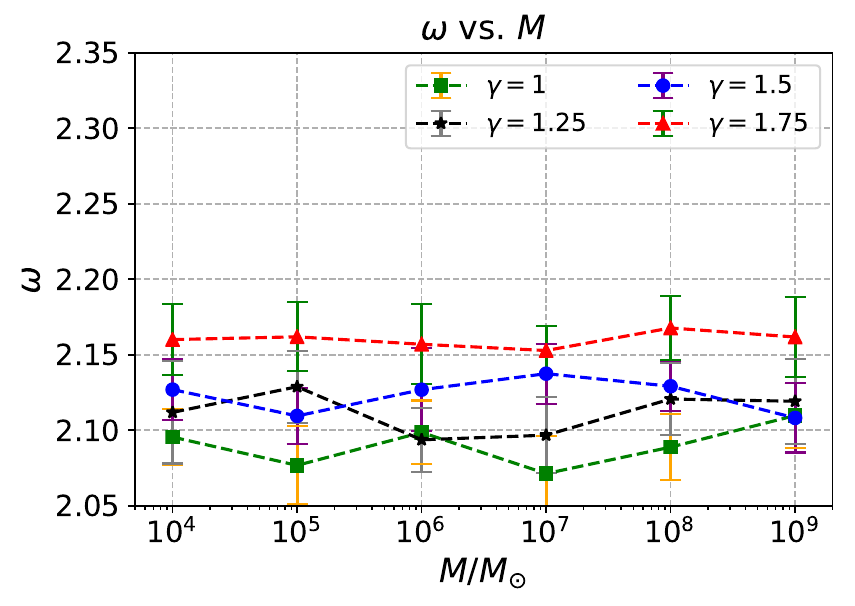}
\includegraphics[width=0.48\textwidth,height = 0.38\textwidth]{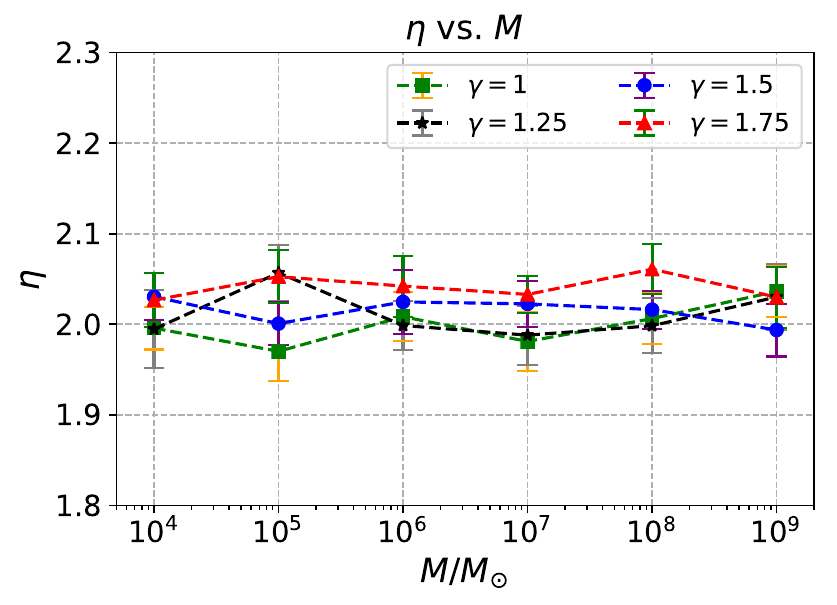}
\caption{$\omega$ and $\eta$ for initial power-law profiles with $\gamma=1,1.25,1.5$ and $1.75$. The green dashed line, blue dashed line, black dashed line, and red dashed line are the results of power-law profiles with $\gamma=1,1.25,1.5$ and $1.75$ respectively. (Left) Numerical scatters of $\omega$ vs. black hole mass $M$. (Right) Numerical scatters of $\eta$ vs. black hole mass $M$.\label{fig:omega and eta1}}
\end{figure*}

After numerically calculating the dark matter spike after adiabatic growth, we fit with the analytic expression below,
\begin{equation}\label{dark matter density}
\rho(r) = \frac{\kappa(\rm{GeV\cdot cm^{-3}})}{(r/GM)^\omega}\left(1-\frac{4GM}{r}\right)^\eta.
\end{equation}
Here $\kappa$ represents the characteristic density of the spike, $\omega$ is the power-law index of distance $r$, and index $\eta$ adjusts the profile approaching the black hole. 
If necessary, the unit can be converted by $\text{GeV}\cdot \text{cm}^{-3} = 0.0265 {M_\odot} \cdot \text{pc}^{-3}$.

We visualize $\kappa$ in Fig.~\ref{kappa} to compare the Hernquist profile and power-law profile with $\gamma=1$. 
The results show that $\kappa$ varies significantly with $M$, while nearly equal for both profiles. 
Then we can use a logarithmic formula to fit the correlation between $\kappa$ and $M$, 
\begin{equation}\label{lg-lg}
\rm{log}\left(\kappa/(GeV\cdot cm^{-3})\right) = \mathit{a}\cdot log(\mathit{M/M_{\odot}})+\mathit{b}.
\end{equation}
The best fit parameters of the initial Hernquist halo and initial power-law profile with $\gamma=1$ are $(a,b)=(-1.616 \pm0.003, 31.37\pm0.02)$ and $(-1.612\pm0.005, 31.35\pm0.03)$, respectively. 
Fig.~\ref{omega and eta} shows the correlation between $\omega$ ($\eta$) and $M$. 
Given that $\omega$ and $\eta$ are nearly constant, we use the average values $(\bar{\omega}, \bar{\eta})$ to simplify the spikes generated by different initial dark halos. 
For the above two initial halos, $(\bar{\omega}, \bar{\eta})$ are $(2.09\pm0.01,1.99\pm0.02)$ and $(2.09\pm0.01,2.00\pm0.02)$, respectively. 
We find that the results $(\kappa, \bar{\omega}, \bar{\eta})$ from the two initial inputs are quite similar. 
It can be explained by the similar behaviors $r^{-1}$ close to the galactic center of two initial profiles. 
Therefore, we can assume that the spikes that grow from the Hernquist profile and the power-law profile with $\gamma=1$ are equivalent near black holes. 

Furthermore, we compute the spikes that grow from power-law profiles with various indices $\gamma=1,1.25,1.5$ and $1.75$, which are motivated by other physical models. 
In Fig.~\ref{fig:kappa1} we show the logarithmic relation between characteristic density of spike $\kappa$ and black hole mass $M$. 
For larger $\gamma$, the density of the spike is higher since the overall density close to the black hole is larger. 
In Fig.~\ref{fig:omega and eta1} we show the numerical fitting results of $(\omega, \eta)$ generated from different initial halos.  
The best fit parameters $(a, b)$ and $(\bar{\omega},\bar{\eta})$ for $\gamma=1,1.25,1.5$ and $1.75$ are tabulated in Tabel.~\ref{tab:fit parameter}. 
It can be found that under conditions of initial power-law profiles, the index $\bar{\omega}$ of spike increases with $\gamma$, but the variation tendency of $\bar{\omega}-\gamma$ is different from $\alpha_\text{sp}=(9-2\gamma)/(4-\gamma)$ given by~\cite{PhysRevLett.83.1719}.
This might be due to the range of distances we investigate is closer to the black hole and we have employed relativistic analysis.

Combining Eq.~\ref{dark matter density} and Eq.~\ref{lg-lg}, we can relate the density of the spike to the mass of the black hole,
\begin{equation}\label{fit density}
\rho(r) = 10^{b} \cdot \left(\frac{M}{M_\odot}\right)^{a} \cdot \left(\frac{r}{GM}\right)^{-\bar{\omega}} \cdot \left(1-\frac{4GM}{r}\right)^{\bar{\eta}} \rm{GeV\cdot cm^{-3}}.
\end{equation}
From the numerical results of $(\bar{\omega},\bar{\eta})$, we can observe the profiles of spikes generated by the above initial halos are quite similar, namely $\bar{\omega}\simeq 2.10$ and $\bar{\eta}\simeq 2.00$. Thus, in the next section, Sec.~\ref{sec:velocity}, without loss of generality, we shall only focus on the dark matter spikes growing from initial Hernquist profiles and analyze the corresponding velocity distribution of dark matter particles. 

\begin{table*}[t]
\caption{\label{tab:fit parameter}%
The best fit parameters $(a, b)$ and $(\bar{\omega},\bar{\eta})$ of initial halos surrounding black holes with mass range from $10^4 M_\odot$ to $10^9 M_\odot$.}
\begin{ruledtabular}
\begin{tabular}{ccc}
Initial halo &$(a,b)$& $(\bar{\omega},\bar{\eta})$\\ \hline
 Hernquist & $(-1.616 \pm0.003, 31.37\pm0.02)$& $(2.09\pm0.01,1.99\pm0.02)$ \\
 Power law $(\gamma=1)$ & $(-1.612\pm0.005, 31.35\pm0.03)$& $(2.09\pm0.01,2.00\pm0.02)$ \\
 Power law $(\gamma=1.25)$ & $(-1.642\pm0.005, 32.05\pm0.03)$& $(2.11\pm0.01,2.01\pm0.02)$ \\
 Power law $(\gamma=1.5)$ & $(-1.677\pm0.004, 32.81\pm0.03)$& $(2.13\pm0.01,2.01\pm0.01)$ \\
 Power law $(\gamma=1.75)$ & $(-1.714\pm0.002, 33.70\pm0.02)$& $(2.16\pm0.04,2.04\pm0.01)$ \\
\end{tabular}
\end{ruledtabular}
\end{table*}

\section{Velocity distribution of dark matter} \label{sec:velocity}

Once we know the density profile of a dark matter profile, next we shall evaluate the velocity distribution of dark matter particles in the spike since it can affect many physical observables. 
In this section, we shall analyze the range of dark matter velocity at different distances from black hole, and fit the velocity distribution of dark matter. We have performed a fit of the Gaussian distribution function in the previous work~\cite{PhysRevD.110.103008}. Nevertheless, the range of dark matter velocity is not arbitrary. 
The lower bound of velocity in Gaussian distribution is inconvenient to define. 
To approximate the velocity distribution range predicted by our analysis and to better normalize the distribution function numerically, we use a formula similar to the Maxwell-Boltzmann distribution, as it provides a better fit near the lower bound of velocity. 

\begin{figure*}[t]
\centering
\includegraphics[width=0.48\textwidth,height = 0.38\textwidth]{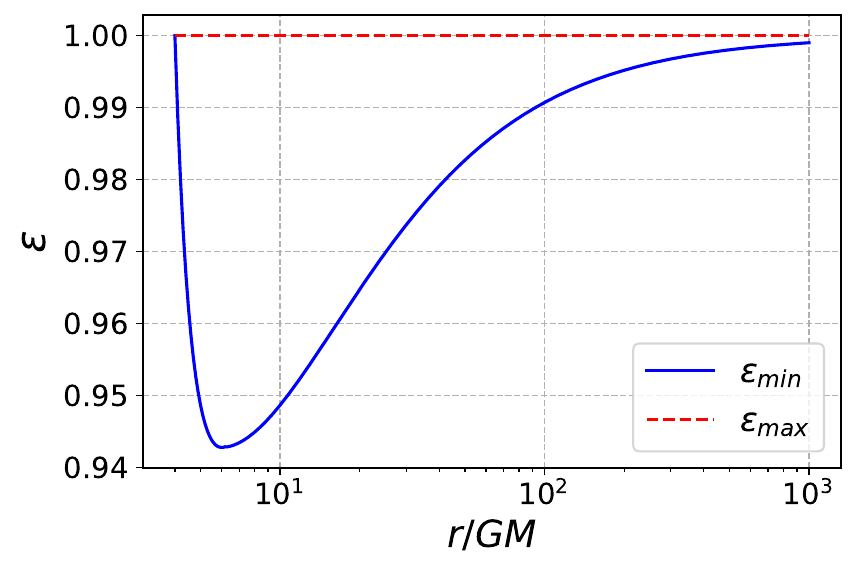}
\includegraphics[width=0.48\textwidth,height = 0.38\textwidth]{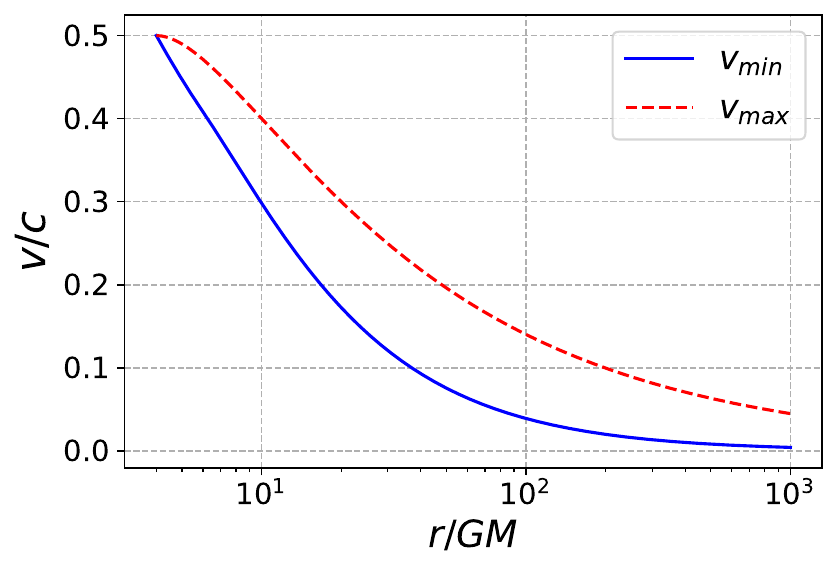}
\caption{The boundaries of relative energy $\varepsilon$ (left) and velocity of dark matter in spike $v$ (right). Both vary with distance $r$. At $r=4GM$ we have $v_\text{min}=v_\text{max}=c/2.$\label{e and v}}
\end{figure*}

In Schwarzschild spacetime, the four-velocity can be written as $u^\mu = \gamma(1,v^j)$ and relative energy per unit mass $\varepsilon = u_{t}=g_{tt}u^t$, where $v^j=u^j/u^t$, $\gamma = (-g_{tt} - g_{ij}v^{i}v^{j})^{-1/2}$ is Lorentz factor, and $g_{\mu\nu}$ is the matrix elements of Schwarzschild metric.
Here we set velocity $v = \sqrt{g_{ij}v^{i}v^{j}}$, then we can express the velocity of dark matter particle by $\varepsilon$,
\begin{equation}\label{v-e}
v = \sqrt{-g_{tt} - \frac{(g_{tt})^2}{\varepsilon^2}},
\end{equation}
where $g_{tt}=-1+2GM/r$. 
It is a monotonic function for $\varepsilon$ when $g_{tt}$ is determined by $r$. 
On the one hand, given $\varepsilon_\text{max}=1$, $v_\text{max}$ can be obtained at various distances.
On the other hand, the minimum values of $\varepsilon$ at different distances are analytic \cite{PhysRevD.88.063522},
\begin{equation}\label{e-max-min}
\varepsilon_\text{min}(r)=\left\{\begin{aligned}
&\frac{1-2GM/r}{\sqrt{1-3GM/r}}, 4GM\leq r \leq 6GM \\
&\frac{1+2GM/r}{\sqrt{1+6GM/r}}, r \geq 6GM 
\end{aligned}\right.,
\end{equation}
thus $v_\text{min}(r)$ also can be expressed by combining Eq.~\ref{v-e} and Eq.~\ref{e-max-min}.
As shown in Fig.~\ref{e and v}, it implies that the velocity of dark matter in a spike is not arbitrary but confined to a range that varies with distance. 
It also conveys that the dark matter particles would escape when $v>v_\text{max}$ or have fallen into BH when $v<v_\text{min}$.
The remaining dark matter contributes to the spike.

\begin{figure*}[t]
\centering
\includegraphics[width=0.48\textwidth,height=0.38\textwidth]{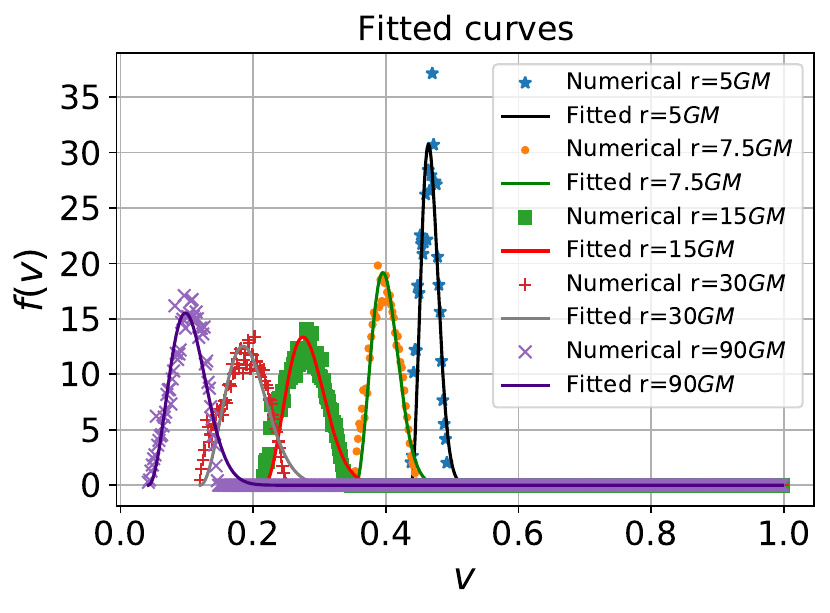}
\includegraphics[width=0.48\textwidth,height=0.38\textwidth]{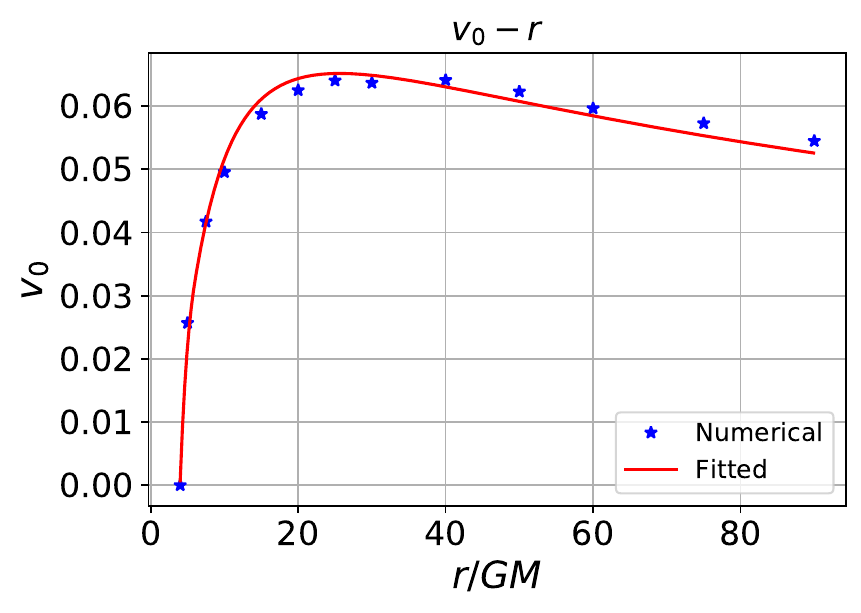}
\caption{The results of numerical scatters and fitted curves. (Left) Velocity distribution function $f(v)$ at various distance $r$. (Right) Characteristic velocity $v_0$ vs. $r$.}\label{distribution of v} 
\end{figure*}

Since the spacetime and density profile are spherically symmetric, we expect the corresponding velocity distribution is also spherically symmetric and isotropic.
Numerically we can obtain the distribution function by differentiating $\rho(r,v)$
\begin{equation}\label{differentiate rho}
f_\text{num}(r,v)=\frac{1}{\rho(r)}\frac{d\rho}{dv}.
\end{equation}
Here we use Maxwell-Boltzmann distribution to fit the results,
\begin{equation}\label{MB}
f_r(v)=\frac{4\pi \left[ v-v_\text{min}(r)\right]^2}{\left[\pi v_0^2(r)\right]^{\frac{3}{2}}} e^{-\left[v-v_\text{min}(r)\right]^2/{v_0^2(r)}},
\end{equation}
where $v_0(r)$ is the characteristic velocity to be fitted with $r$.
It is normalized by 
\begin{equation}\label{normalize}
\int_{v_\text{min}(r)}^{v_\text{max}(r)}f_r(v)dv=1.
\end{equation}
Using the Maxwell-Boltzmann distribution is convenient to change the lower limit of velocity when integrating the distribution function, and it can reflect the Gaussian behavior and the velocity volume element $4\pi v^2dv=d^3v$ as shown in the left plot of Fig.~\ref{distribution of v}.
When fitting the distribution function at $r$, we find that $v_0$ is close to $(v_\text{max}-v_\text{min})/2$. In Fig.~\ref{distribution of v}, we show the velocity distribution at several distances in the left panel and $v_0$ from $4GM$ to $90GM$ in the right panel. We can see that $f_r(v)$ is sharper when getting closer to the black hole. At $r\rightarrow 4GM$, $v_0\rightarrow0$ and the distribution function is approximately a $\delta$-function. This can be understood as the consequence of the smaller phase space for dark matter particles near the black hole.

With the above results, we can further extend Eq.~\ref{fit density} by relating the velocity distribution $f_r(v)$, 
\begin{equation}\label{velocity-density}
\rho(r,v)=\rho(r)\int_{v_\text{min}(r)}^{v}f_r(v')dv'.
\end{equation}
With velocity distribution in hand, in the next section, we shall consider dark matter self-annihilation, and investigate the effects of velocity distribution on the annihilation cross section and the relic density of the spike.

\section{Dark matter Profile with Self-annihilation}\label{sec:anni}
We now consider dark matter self-annihilation and estimate the relic density of the dark matter spike near galactic center. 
If dark matter annihilates, its density is possible to be different from the one we obtain in Sec.~\ref{sec:dark matter}. 

We define $m_\chi$ as the rest mass of the dark matter particle $\chi$, $\left \langle \sigma v \right \rangle$ as the annihilation cross section, and $t_\text{BH}$ as the age of black hole. 
In our calculation, we set $t_\text{BH}=10\text{Gyr}$.  
Denoting the number density as $n$ and energy density $\rho=nm_\chi$, combining the annihilation rate $\Gamma=n\left \langle \sigma v \right \rangle$ and change rate of particle number $dn/dt=-n\Gamma$, we obtain 
\begin{equation}\label{drho/dt}
\frac{d\rho}{dt}=m_\chi \frac{dn}{dt}=-\frac{\rho^2 \left \langle \sigma v \right \rangle}{m_{\chi}}.
\end{equation}
Then we set the spike density in Sec.~\ref{sec:dark matter} as the initial density $\rho_\text{ini}(r)$. 
After the annihilation time $t_\text{BH}$, the final relic density of the dark matter spike is $\rho_\text{relic}(r)$. 
This process can be described by integrating Eq.~\ref{drho/dt}
\begin{equation}\label{int rho}
\int_{\rho_\text{ini}(r)}^{\rho_\text{relic}(r)} -\frac{m_\chi d\rho}{\rho^2 \left \langle \sigma v \right \rangle}=\int_0^{t_\text{BH}} dt = t_\text{BH}.
\end{equation}
For compactness, we can introduce the following constant density,
\begin{equation}\label{rho_core}
\rho_\text{core}=\frac{m_{\chi}}{\left \langle \sigma v \right \rangle t_\text{BH}},
\end{equation}
and solve the relic density from Eq.~\ref{int rho} at $t_\text{BH}$
\begin{equation}\label{relic density}
\rho_\text{relic}(r)=\frac{\rho_\text{core}\rho_\text{ini}(r)}{\rho_\text{core} + \rho_\text{ini}(r)}.
\end{equation}
We have obtained $\rho_\text{ini}(r)$ in Sec.\ref{sec:dark matter}. $\rho_\text{core}$, on the other hand, is determined by the details of the annihilation cross section $\left \langle \sigma v \right \rangle$.

As an illustration, we only consider the tree-level annihilation process of dark matter $\chi$ into the lighter particle $\phi$, $\bar{\chi} \chi \rightarrow \phi \phi$. 
The expressions of annihilation cross section are 
\begin{equation}\label{anniflux}
(\sigma v_\text{rel})_\text{anni}=\left\{\begin{aligned}
&\frac{\pi \alpha_\chi^2}{m_\chi^2}\sqrt{1-\frac{m_\phi^2}{m_\chi^2}},\; \text{s-wave} \\
&\frac{3\pi \alpha_\chi^2}{4m_\chi^2}v_\text{rel}^2\sqrt{1-\frac{m_\phi^2}{m_\chi^2}},\; \text{p-wave} 
\end{aligned}\right.,
\end{equation}
where $\alpha_\chi$ is the dark fine structure constant, $m_\phi$ is the mass of the mediator which generates s-wave and p-wave as scalar and vector particles respectively, and $v_\text{rel}=|\vec{v}_1-\vec{v}_2|$ is the relative velocity of two dark matter particles.
For the s-wave case, the upper line of Eq.~\ref{anniflux} can be directly utilized. 
However, for the p-wave case, the annihilation cross section depends on the velocity.
\begin{figure*}[t]
\centering
\includegraphics[width=5in]{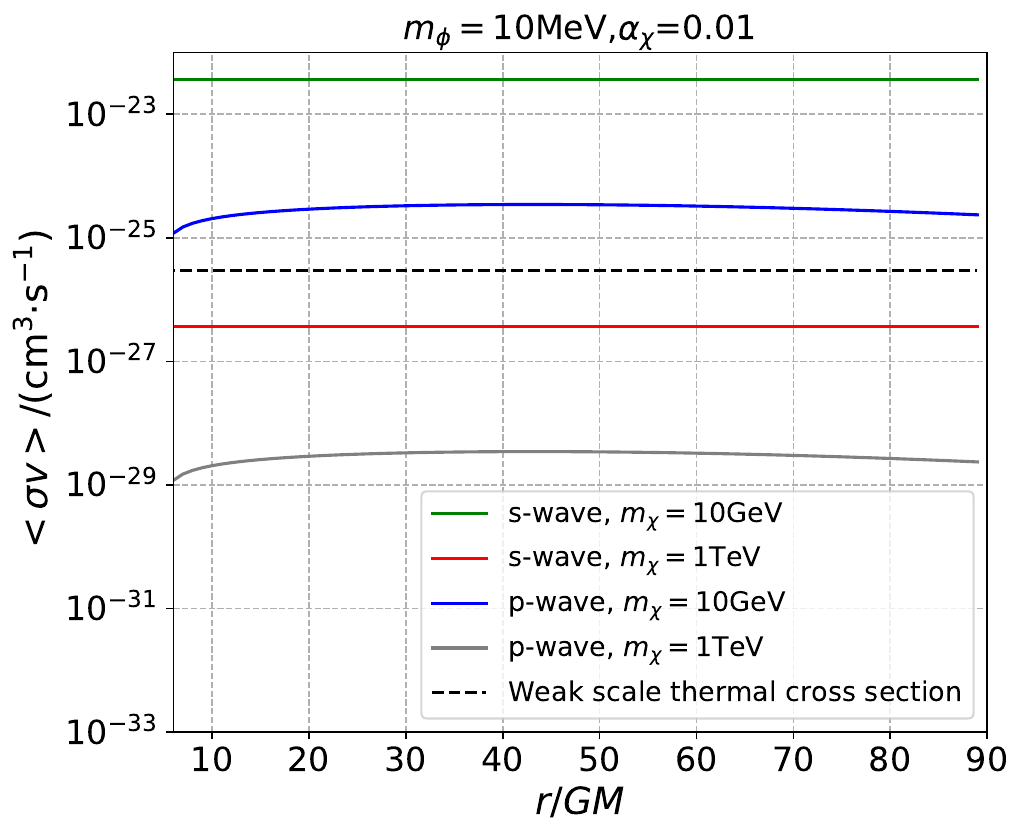}
\caption{Velocity-averaged annihilation cross section varying with distance $r$. 
The mass of the black hole is $5\times10^6 M_{\odot}$. 
The black dashed curve represents the cross section of weak scale thermal relic.
The green and blue solid curves are respectively s-wave cross section and p-wave cross section of the model $(m_\chi,m_\phi,\alpha_\chi)$=(10GeV, 10MeV, 0.01).
The red and grey solid curves are respectively s-wave cross section and p-wave cross section of the model $(m_\chi,m_\phi,\alpha_\chi)$=(1TeV, 10MeV, 0.01).\label{flux} }
\end{figure*}

\begin{figure*}[t]
\centering
\includegraphics[width=5in]{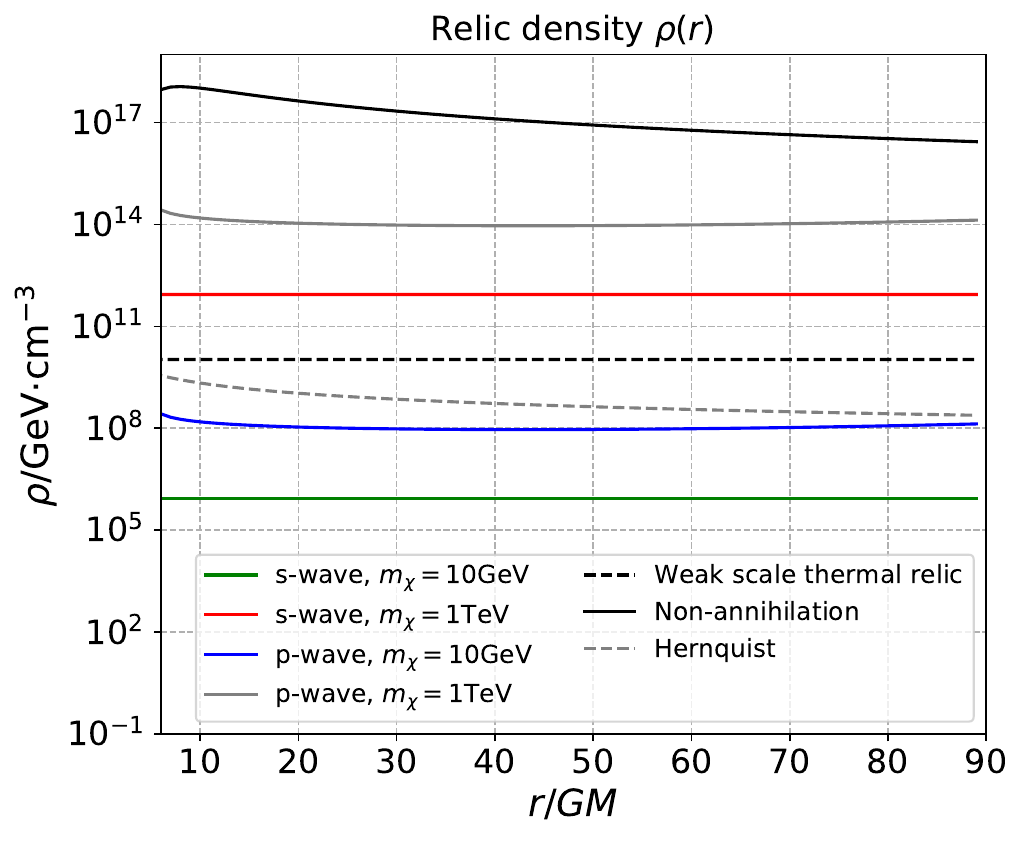}
\caption{Relic density of dark matter varies with distance $r$. The mass of the black hole is $5\times10^6 M_{\odot}$. 
The grey dashed curve represents the initial Hernquist profile of the black hole,
the black solid  curve is for the spike without dark matter self-annihilation,
the black dashed curve for the relic density after weak scale thermal relic, the green and blue solid curves for the relic density after s-wave annihilation and p-wave annihilation of the model $(m_\chi,m_\phi,\alpha_\chi)$=(10GeV, 10MeV, 0.01), respectively, and
the red and grey solid curves for the relic density after s-wave annihilation and p-wave annihilation of the model $(m_\chi,m_\phi,\alpha_\chi)$=(1TeV, 10MeV, 0.01).\label{Relic density} }
\end{figure*}
Therefore, we should estimate the velocity-averaged cross section by Eq.~\ref{MB}. 
The velocity-averaged cross section can be written as, 
\begin{equation}\label{velocity-averaged flux}
\left \langle \sigma v_\text{rel} \right \rangle_\text{anni}(r)=
\int dv_1 dv_2 f_r(v_1)f_r(v_2) (\sigma v_\text{rel})_\text{anni}=\int dv_\text{rel} \frac{4\pi v_\text{rel}^2}{\left[2\pi v_0^2(r)\right]^{\frac{3}{2}}}e^{-\frac{v_\text{rel}^2}{2v_0^2(r)}}(\sigma v_\text{rel})_\text{anni}.
\end{equation}
We consider two groups of dark matter parameters $(m_\chi,m_\phi,\alpha_\chi)$=(10GeV, 10MeV, 0.01) and $(m_\chi,m_\phi,\alpha_\chi)$=(1TeV, 10MeV, 0.01), in comparison with the weak-scale thermal cross section, $\left \langle \sigma v \right \rangle = 3\times 10^{-26}\rm{cm^3\cdot s^{-1}}$ and $m_{\chi}$ = 100GeV.
In Fig.~\ref{flux} and Fig.~\ref{Relic density}, we demonstrate variations of several annihilation cross sections with distance $r$, and relic densities under different dark matter models. 
It is observed that the annihilation cross sections of s-wave and weak scale thermal relic are constant. 
Since the velocity distribution $f_r(v)$ depends on distance, the p-wave annihilation cross sections and relic densities would also vary with distance. 
Near the black hole, the density of spike $\rho_\text{relic}$ is limited to $\rho_\text{core}$. 
$\rho_\text{core}$ is approximately proportional to $m_{\chi}^3$. 
Therefore, if $m_\chi$ is larger, $\rho_\text{relic}$ shall be higher but not more than $\rho_\text{ini}(r)$. 

Now we qualitatively analyze the variation of $\rho_\text{relic}(r)$ when black hole mass changes. 
Eq.~\ref{fit density} and Table.~\ref{tab:fit parameter} show that $\rho_\text{ini}(r)$ is negatively correlated to black hole mass. 
If we consider the situation of a more massive black hole, $\rho_\text{ini}(r)$ might be lower than $\rho_\text{core}$. 
Eq.~\ref{relic density} implies that $\rho_\text{relic}(r) < \text{min}\{\rho_\text{core}, \rho_\text{ini}(r) \}$. 
This means that the density of dark matter spikes after annihilation is approximately the smaller one in $\rho_\text{core}$ and $\rho_\text{ini}(r)$. 
Therefore, around more massive black holes, the density profile is almost $\rho_\text{ini}(r)$, a spiky structure. 
In contrast, for smaller black holes, $\rho_\text{core}$ will dominate, leading to a flat profile.

\section{Discussion} \label{sec:discussion}
Through numerical methods, we obtained the density profiles of a series of dark matter spikes around Schwarzschild black holes of various masses. However, our results differ from those of the attractor solutions~\cite{10.1111/j.13652966.2011.18687.x, 10.1111/j.13652966.2011.19258.x}. In this section, we conduct a comparative analysis between dark matter spike and Bondi/Michel solutions (two special cases of the attractor solution) to clarify these differences and mitigate potential confusion.

The two methods differ fundamentally in their treatment of dark matter. The dark matter spike considers collisionless cold dark matter particles, whereas the Bondi/Michel solutions treat dark matter as an ideal fluid.

The dynamical frameworks governing two methods are also distinct. Dark matter spike is generated via adiabatic contraction, driven by the strong gravitational potential of the central black hole. During this process, dark matter particles undergo gradual orbital readjustment and energy redistribution, leading to a steepened central density profile.  
The dynamics is influenced by the gravity of the black hole and the orbital evolution of dark matter particles, with black hole accretion typically neglected in this scenario. 
In contrast, Bondi/Michel solutions describe the spherically symmetric accretion of fluids in the Newtonian gravitational field of a black hole (Bondi solution) or the accretion with relativistic corrections (Michel solution). 
These solutions characterize the equilibrium density distribution of accreted dark matter, with dynamics dictated by the hydrodynamic continuity equation, Euler equation, and the dark matter equation of state.
Here, the accretion of black hole constitutes the dominant physical mechanism.

Accordingly, the two situations require different boundary conditions. For dark matter spike, at far field, the profile matches the initial halo on account of the weakening of the gravitational potential of black hole. For Bondi/Michel solutions, the density and the speed of sound at infinity should be given. In Schwarzschild spacetime, dark matter spike vanishes near the black hole $r\leq4GM$ due to the absence of stable orbital phase space. 
Bondi/Michel solutions, however, typically extend inward to $r= 1GM$, resulting in markedly distinct profiles between the different scenarios.
The velocity distribution of dark matter spike is isotropic and spherically symmetric, depending solely on the velocity magnitude $v$ and the radial distance $r$ from the black hole. In contrast, Bondi/Michel solutions for spherically symmetric accretion exhibit a velocity distribution determined exclusively by the radial velocity component $v_r$.
\textcolor{red}{An example comparison is presented in the Appendix.~\ref{app:camparison}.}

Finally, the evolution durations and stability of the two densities are determined by different factors. The formation of dark matter spike takes a long time scale close to the age of the galaxy (about $10~\text{Gyr}$). And the profile might be disrupted by subsequent dynamical processes, such as stellar scattering, black hole mergers, and the self-interaction of dark matter as mentioned in Sec.~\ref{sec:dark matter}. In addition, as discussed in Sec.~\ref{sec:anni}, spikes with higher density are probable to be reduced because of the annihilation of dark matter. In other words, the stability of dark matter spike depends on the local galactic environment and additional assumptions of steady state. 
Bondi/Michel solutions involve the preconditions of stationary accretion. 
And the time scale of evolution is relatively short ($10^5-10^7~\text{yr}$). 
Consequently, the obtained profiles are stable.

The conclusion is that both methods are theoretical models involving the distribution of matter around compact celestial bodies, but the physical backgrounds, mathematical descriptions and application scenarios are quite different. The determination and distinction of dark matter profiles around central black holes would require future observational efforts.

\section{SUMMARY} \label{sec:CON}
We have employed the relativistic framework to compute the density of dark matter spikes around Schwarzschild black holes that range from $10^4 M_{\odot}-10^9 M_{\odot}$. 
By correlating the black hole mass with initial dark halo parameters, we have established the relation between the parameters of the inner dark matter spike and the black hole mass with a fitting formula Eq.~\ref{fit density}. For the initial Hernquist halo and power-law halo, we have used a series of parameters to describe dark matter spikes, which can simplify future phenomenological studies with relativistic spikes.

We have also estimated the velocity distribution of dark matter particles in the density spike. 
As an application, we have applied it to the p-wave annihilation cross section near the black hole and calculated the relic density of the dark matter spike in galactic center. 
Compared to the s-wave case, the p-wave annihilation cross section is suppressed by the velocity. By numerical and qualitative analysis, we have demonstrated that, for dark matter with a p-wave annihilation cross section, the spike shall be less dissipated and the relic density will not be a constant at the inner halo. In addition, the spike is less affected by annihilation around young and less massive black holes. We expect these results might provide valuable insights for future multi-messenger dark matter detections.

\begin{acknowledgments}
This work is supported by the National Key Research and Development Program of China (Grant No.2021YFC2201901), the National Natural Science Foundation of China (Grant No.12347103) and the Fundamental Research Funds for the Central Universities. 
We acknowledge the use of \texttt{NumPy} \cite{Harris_2020}, \texttt{SciPy} \cite{Virtanen_2020}, and \texttt{Matplotlib} \cite{4160265} for numerical calculations and data visualization.
\end{acknowledgments}

\appendix
\section{The stability of dark matter spike}
\label{app:stability}
{In our framework, we consider the process by which an initial dark halo forms a steeper spike through adiabatic growth under the gravitational potential of a black hole. However, an astrophysical model should be stationary to ensure the observability.
We neglect the astrophysical mechanisms that might disrupt the spike. Moreover, what we aim to explain here is whether the model is self-consistently stable after formation. 
We will qualitatively discuss the stability of adiabatic spike. } 

\subsection{The stability reflected by the mass current four-vector}
{To begin with, we need to simplify the Eq.~\ref{mass current}. 
Using the notation of our pervious work~\cite{PhysRevD.110.103008},
the four constants of motion in Schwarzschild spacetime are
\begin{equation}\label{constants_motion}
\begin{aligned}
 \varepsilon &\equiv -u_t = -g_{tt}u^t, \;\mu = \sqrt{-p_\mu p^\mu}, \\
 L^2 &\equiv (g_{\theta\theta}u^\theta)^2 + \frac{L_z^2}{\sin^2\theta}=(u_\theta)^2 + \frac{(u_\phi)^2}{\sin^2\theta},\\
L_z &\equiv u_\phi=g_{\phi\phi}u^\phi.
\end{aligned}
\end{equation}
Utilizing the relation $d^4p=|J|^{-1}d\varepsilon dL^2dL_zd\mu$, we transform the integral volume element. Here the Jacobian is 
\begin{equation}\label{jacobian}
\begin{aligned}
 J &\equiv \left| \frac{\partial(\varepsilon,L^2,L_z,\mu)}{\partial(p^{t},p^{r},p^{\theta},p^{\phi})} \right| = -2\mu^{-3}r^4(r^2-2GMr)u_ru^{\theta}\sin^2\theta.
\end{aligned}
\end{equation}
Taking into account the distribution function $f^{(4)}(x,p) = \mu^{-3}f(\varepsilon,L^2,L_z)\delta(\mu-\mu_0)$, we obtain 
\begin{equation}\label{masscurrent1}
J^\mu(r)=\iiint d\varepsilon dL^2dL_z \frac{u^{\mu}f(\varepsilon,L^2,L_z)}{2r^4(r^2-2GMr)u_ru^{\theta}\sin^2\theta}.
\end{equation}
Considering the spherical symmetry of Schwarzschild spacetime, we find that the components $J^r,J^\theta$ and $J^\phi$ are zero after integrating in the total phase space. 
Therefore, the nonzero component of the mass current four-vector is only $J^t$, which has been used in our calculation. 
The above fact conveys that $J^\mu=(J^t,0,0,0)$. Combining the definition $J^\mu=\rho u^\mu$, we have $u^\mu=(u^t,0,0,0)$. 
It implies that as we consider all the dark matter particles around a static Schwarzschild black hole as a relativistic fluid, the fluid macroscopically appears stationary in the spacetime. 
Thus, the spike in the vicinity of black hole is stationary.
}

\subsection{The gravitational wave emission of dark matter particles}
{In the background dominated by the gravitational potential of the black hole, the gravitational radiation from orbital energy loss and the gravitational drag of the surrounding material should be the main pathways for the evolution of dark matter particles in spike. We consider the situation close to the black hole, where gravitational radiation dominates. 
We have implemented calculation for the gravitational wave of extreme mass-ratio inspiral in Fig.~5 of  previous work~\cite{PhysRevD.110.103008}. 
For a compact object with the mass of $10M_\odot$ starts to orbit the galactic black hole from about $r=10GM$, the time to plunge is about $t_{\textrm{plunge}}=10~$ years.
The average orbital energy loss efficiency per unit mass~\cite{blanchet2024postnewtoniantheorygravitationalwaves,10.1093/ptep/ptv092} is 
\begin{equation}\label{energy loss}
\left< \frac{dE}{dt} \right>_\text{DM}\sim m_\text{DM}.
\end{equation}
Accordingly, the time scale of dark matter particles plunging into the galactic black hole can be estimated as
\begin{equation}\label{plunge}
t_\text{DM}\sim 10\times \frac{10M_\odot}{m_\text{DM}} \text{years}.
\end{equation}
For $m_\text{DM}=100\text{GeV}$, $t_\text{DM}\approx10^{59}~\text{years}$. 
Hence, without additional astrophysical processes, such dark matter spikes can be considered stationary within the age of the universe. 
}

\subsection{Relaxation time}
{The formation of dark matter spikes during the process from adiabatic growth of the initial steady-state distribution to the current distribution corresponds to the two-body relaxation timescale. 
To ensure the stability, we require the age of spike to be older than the simplified relaxation time~\cite{sigurdsson2003adiabaticgrowthmassiveblack,Binney2008GalacticDS,Merritt+2013,Bar-Or_2019}: 
\begin{equation}
t_\text{relax}\approx \frac{0.1}{\text{ln}\Lambda}(\frac{r^3}{GM_{\text{BH}}})^{\frac{1}{2}},
\end{equation}
where $\text{ln}\Lambda \sim O(1)$ is Coulomb logarithm. 
Take the galactic black hole ($M_\text{BH}\approx 4\times 10^6 M_\odot$) into account, 
within the distance $r<0.01\text{pc}$, 
$t_\text{relax} < 10^7 \text{years}$ is shorter than Hubble time. 
This means that the dark matter particles close to the center have undergone a sufficient period of time to approach dynamical equilibrium. 
Therefore, in the region near the black hole we consider, the dark matter spike should be stationary. 
}

\section{The comparison between attractor solution and dark matter spike}
\label{app:camparison}
{Here we conduct a specific comparison to demonstrate the differences between attractor solution~\cite{10.1111/j.13652966.2011.18687.x} and dark matter spike in density and velocity profiles. 
We show their different results under the gravitational effect of massive black hole.

We respectively consider the formation of a dark matter spike and a stable dark matter accretion flow around a Schwarzschild black hole with the mass of $10^6 M_\odot$. 
For dark matter spike, we generate the corresponding profile based on the results of Table.~\ref{tab:fit parameter} and Eq.~\ref{fit density}. The spike grows from an initial Hernquist halo. 
For attractor solution, in addition to the initial density of the dark matter fluid, the equation of state of dark matter is also an important precondition, which is described by
\begin{equation}\label{EoS}
p=(\Gamma-1)\epsilon \rho.
\end{equation}
Here, $\Gamma$ is the adiabatic index that reflects the self-interacting intensity of dark matter, and $\epsilon$ is the specific internal energy, $p$ is the pressure and $\rho$ is the rest mass density of dark matter fluid. 

We adopt the steady-state density and velocity profiles that are shown in Fig.~5 of Ref.~\cite{10.1111/j.13652966.2011.18687.x}, where $\Gamma=1.1$, $\epsilon=0.5$ and the initial velocity of the fluid is $v_r=-0.1$. The negative velocity represents that the fluid flows towards the black hole. 
According to the results presented in the above reference, the initial density solely influences the magnitude of the steady-state density while leaving its spatial profile unchanged, provided that the dark matter parameters $\Gamma$ and $\epsilon$ remain constant. 
Therefore, we select the initial density as $\rho_0=10^{-12}$ in geometric unit $1/M^2_\text{BH}$, corresponding to $3.5\times 10^{17} \text{GeV}\cdot \text{cm}^{-3}$, to adjust the density to align with that of the spike and visually compare the differences of the profiles. 
The density and velocity profiles are exhibited in Fig.~\ref{att_vs_spike} and Fig.~\ref{v_att_vs_spike}, respectively.

\begin{figure*}[t]
\centering
\includegraphics[width=5in]{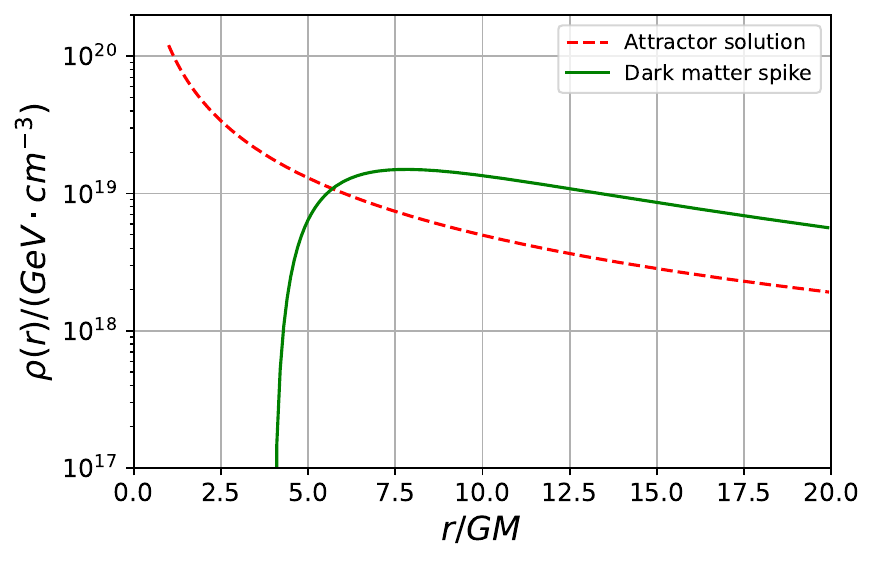}
\caption{Density profiles of dark matter spike and attractor. The mass of black hole is $10^6 M_\odot$. The dark matter spike (green solid curve) grows from a initial Hernquist halo. The initial parameters are determined by mass-velocity-dispersion relation introduced in Sec.~\ref{sec:dark matter}. The attractor solution (red dashed curve) is the steady-state density of dark matter accreted by the black hole. The parameters in the equation of state Eq.~\ref{EoS} are $(\Gamma,\epsilon)=(1.1,0.5)$. The initial density of dark matter is $\rho_0=10^{-12}/M^2_\text{BH}=3.5\times 10^{17} \text{GeV}\cdot \text{cm}^{-3}$. \label{att_vs_spike}}
\end{figure*}

As shown in Fig.~\ref{att_vs_spike}, 
the dark matter spike exhibits a steep, cuspy density profile, decreases as $r^{-\gamma}$ with $\gamma>2$ at large radii, characteristic of regions where dark matter is highly concentrated due to gravitational interactions with a central black hole. 
Since the absence of stable orbital phase space, the profile is cut off at $r=4GM$. 
The result is distinct with the earlier work~\cite{1989ApJ...336..313P} treated by spherically symmetric accretion of Schwarzschild black hole, in which the power-law index of dark matter fluid was close to the range of $5/3<\gamma<2$.
In contrast, the attractor solution shows a gradual and flattened density profile as $\sim r^{-\gamma}$ with $\gamma<1.5$. 
The curve stops at $r=GM$, closer to black holes than dark matter spike.
It represents a stable equilibrium configuration where the self-interaction of dark matter and the accretion of black hole have smoothed the density distribution. 
Under the initial conditions we chose, near the black hole, the density of the dark matter spike is generally higher than that of the attractor solution. 
Considering that the attractor solution is approximately proportional to the initial density $\rho_0$. For a larger initial density, the result of attractor can be higher than that of dark matter spike. However, when the initial Hernquist halo ($\sim 10^{12} \text{GeV/cm}^3$ near the black hole) in our framework is applied to the initial conditions of the attractor, the steady-state density may be much lower than that of the dark matter spike.
In other words, considering only the same initial density, for the dark matter fluid with pressure in the Ref.~\cite{10.1111/j.13652966.2011.18687.x}, the enhancement in dark matter density caused by the attractor solution resulting from the accretion of black hole is less than that of the spike formed by adiabatic growth.

\begin{figure*}[t]
\centering
\includegraphics[width=5in]{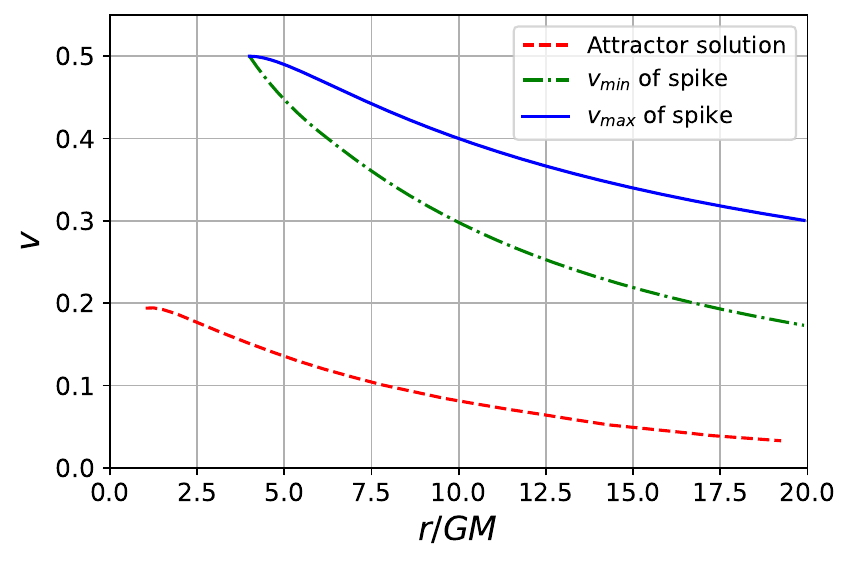}
\caption{Velocity profiles of dark matter spike and attractor. 
The red dashed curve is the velocity field of attractor solution with the initial velocity $|v_r|=0.1$. The initial density $\rho_0=3.5\times 10^{17} \text{GeV}\cdot \text{cm}^{-3}$, the parameters of dark matter $(\Gamma,\epsilon)=(1.1,0.5)$. 
The green dashed curve and blue solid curve are the variations of the minimum and maximum velocities of particles in spike with distance, respectively.
\label{v_att_vs_spike}}
\end{figure*}

We take the absolute values of the results of attractor for comparison, as shown in Fig.~\ref{v_att_vs_spike}. Intuitively, the velocities of dark matter particles in spike are significantly higher than those of the dark matter fluid described by attractor solution, which demonstrates the dynamic difference between the orbital motion of particles and the accreted fluid under the gravitational effect. 
As evidenced by the results in Fig.~\ref{distribution of v}, the dark matter particles within the spike exhibit a diversity of orbital configurations, leading to a velocity distribution and establishing a state of dynamical equilibrium. 
The velocity of the fluid described by attractor at a certain distance is determined, representing the radial velocity of the fluid towards the black hole. Unlike the spike, the velocity profile enters the region near $r=GM$.}

We consider a reference case that, for dark matter fluid accreted by the spherically symmetric black hole, we have the relation at a certain time~\cite{1989ApJ...336..313P},
\begin{equation}\label{balance}
4\pi \rho r^2 v = \textrm{constant}. 
\end{equation}
If the fluid is in free fall velocity $v=\sqrt{2GM/r}$,
so $\rho \sim r^{-3/2}$. 
We notice that the power-law index the above is higher than the profile of attractor in Fig~\ref{att_vs_spike}. 
The reason is that dark matter fluid resists black hole accretion through pressure, reducing the speed of falling into the black hole and leading to a flat density profile. 
On the other hand, the dark matter in spike is a series of particles that resist falling into the black hole through orbital motion. 
Therefore, the velocity boundary (the green dashed and blue solid curves) determined by the stable orbital phase space reflects the velocity range of dark matter particles, and consequently we can investigate the velocity distribution. 
\color{black}

\bibliography{citation}
\bibliographystyle{apsrev}

\end{document}